\documentstyle[12pt,epsfig]{article}
\newcommand{\lessgtr}{\begin{array}{c} < \\[-1.7ex] > \end{array}}
\newcommand{\smalllessgtr}{\!\!\begin{array}{c} \mbox{\tiny $<$} \\[-2.1ex] 
                                            \mbox{\tiny $>$} \end{array}\!\!}

\newcommand{\fig}[1]{fig.\ref{fig:#1}}

\newcommand{\pkt}{\, .}
\newcommand{\komma}{\, ,}

\newcommand{\Res}{\mbox{Res}}

\newcommand{\tila}{\tilde{a}}
\newcommand{\tilb}{\tilde{b}}
\newcommand{\tilc}{\tilde{c}}
\newcommand{\iint}{\int\!\!\!\!\int}

\newcommand{\pvint}{{\mbox{P$\!$.V$\!$.}}\!\int}

\begin{document}

 \thispagestyle{empty}
 \begin{flushright}
 {MZ-TH/96-30}    \\
 {hep-ph/9610285} \\
 {September 1996}           \\
 \end{flushright}
 \vspace{1cm}
 \begin{center}
 {\large \bf
New representation of the two-loop crossed vertex function}
 \end{center}
 \vspace{1cm}
 \begin{center}
 {Alexander Frink, Ulrich Kilian, Dirk Kreimer\footnote{
email: {\em author}@dipmza.physik.uni-mainz.de}}\\
  \vspace{.3cm}
 {\em Institut f\"ur Physik, Johannes Gutenberg-Universit\"at,\\
 \mbox{D-55099} Mainz, Germany}
 \end{center}
 \hspace{3in}
\begin{abstract}
   We calculate the two-loop vertex function for the crossed
topology, and for arbitrary masses and
external momenta. We derive a double integral
representation, suitable for a numerical evaluation
by a Gaussian quadrature.
Real and imaginary parts of the diagram can be calculated separately.
 \end{abstract}

 \vspace{15mm}

 \begin{center} PACS numbers: 02.70.+d, 12.38.Bx, 11.20.Dj
 \end{center}
\newpage


\newcommand{\wquad}{s_0^2+s_0(\sigma_1+\sigma_2)+\sigma_1\sigma_2}
\newcommand{\wsigsig}{\sqrt{\sigma_1\sigma_2}}

\section{Introduction}

In the recent past many efforts have been made to calculate
two-loop Feynman diagrams with masses, e.g. 
\cite{dirk1,dirk2,newrepplanar,tarasov,kato}.
Though now very effective methods exist for the general mass case
of the two-loop two point function \cite{xloops,baub}, 
these methods need considerable
extensions or modifications to cope with two-loop three-point functions
in general.
The planar topology has
been discussed extensively, e.g.~in \cite{dirk2, newrepplanar,
fujimoto92,fleischer93}. For the other important topology --- the crossed
topology --- so far two methods have been presented.

The method presented in \cite{kato} is based on Feynman parameters
and uses high dimensional Monte Carlo integration, resulting
in extensive CPU usage and slow convergence when high accuracy is
needed.

Taylor expansion, analytic continuation and Pad{\'e} approximations,
as presented in \cite{tarasov}, gives results with very high
accuracy, but the method is, at this stage,
still restricted to one kinematical variable.

Our approach is based on \cite{dirk2,newrepplanar}, where it gave
excellent results for the planar two-loop three point function.
The calculation of the crossed topology is similar to the 
planar case. Again we succeed by considering
orthogonal and parallel space variables. From this
starting point, a four-fold
integration is immediately obtained. 
Still following the lines of \cite{newrepplanar}
we use Euler transformations for subsequent integrations.
The difference between the planar and the non-planar topology
is apparent  in the increasing number of different cases which 
have to be considered
for the latter, while, fortunately,  
the conceptual frame remains unaffected. 
Once more we end up with a two-fold integral over a finite region,
solely involving dilogarithms and related functions. This integral
representation allows for a Gaussian quadrature, and is thus
perfectly suitable for practical purposes \cite{FKKR}.

\section{Calculation}
{
\begin{figure}[ht]
\begin{center}
\fbox{
\epsfig{file=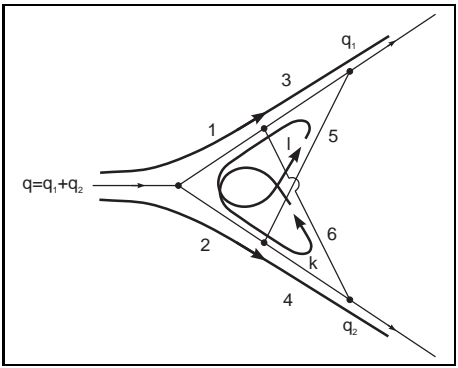,width=5.5cm}
}
\end{center}
\caption {The two-loop crossed vertex function.}
\label{fig:crossedfig}
\end{figure}
}
Our task is to calculate the two-loop crossed vertex function,
as shown in fig.\ref{fig:crossedfig}.
We restrict ourselves to the scalar integral, which is given by
\begin{equation}
\label{eq:theintegral}
V(q_1,q_2) = \int\!\! d^4l \int\!\! d^4k \frac{1}{P_1 P_2 P_3 P_4 P_5 P_6}
\komma
\end{equation}
where $q=q_1+q_2$ is the momentum of the incoming particle
and $q_1$ and $q_2$ are the momenta of the outgoing particles.
Further, $l$ and $k$ are the internal loop momenta, and $P_i$, $i=1 \dots 6$
are the propagators
of the internal particles, as labeled in fig.\ref{fig:crossedfig}.
We assume that all external particles are massive and have time-like
momenta. The massless limit is approached smoothly
as long as no on-shell singularities occur.

The integral will be splitted into parallel and orthogonal space 
integrations. The parallel
space is two-dimensional here, hence leaving two dimensions for the
orthogonal space. The integral is convergent in
$D=4$ dimensions, and there is no need to use a regularization scheme. 

Since the integral is Lorentz invariant, the splitting into
parallel and orthogonal space variables is equivalent
to the choice of a Lorentz frame. For time-like
external particles, we choose the rest frame of the incoming
particle, and assume that outgoing particles are
moving along the $x$-axis. 

Explicitly, the four-momenta are chosen as
\begin{eqnarray}
q^\mu & = & (e_1+e_2,0,0,0)
\nonumber\\
q_1^\mu & = & (e_1,q_z,0,0)
\nonumber\\
q_2^\mu & = & (e_2,-q_z,0,0)
\nonumber\\
l^\mu & = & (l_0,l_1,\vec{l}_\perp)
\nonumber\\
k^\mu & = & (k_0,k_1,\vec{k}_\perp)
\label{eq:momenta}
\end{eqnarray}
with $q_z$, $e_1$ and $e_2$ positive and $q_z<e_1$, $q_z<e_2$.

The propagators $P_i$ of the internal particles can be written down
using the flow of loop momenta as shown in fig.\ref{fig:crossedfig}
\begin{eqnarray}
P_1 & = & (l+k-q_1)^2-m_1^2+i\eta
\nonumber\\
P_2 & = & (l+k+q_2)^2-m_2^2+i\eta
\nonumber\\
P_3 & = & (l-q_1)^2-m_3^2+i\eta
\nonumber\\
P_4 & = & (k+q_2)^2-m_4^2+i\eta
\nonumber\\
P_5 & = & l^2-m_5^2+i\eta
\nonumber\\
P_6 & = & k^2-m_6^2+i\eta \pkt
\end{eqnarray}
A small imaginary part $-i\eta$ with $\eta > 0$ is  assigned to the 
squared masses of the internal
particles and is chosen equal for all particles for convenience.
Hence differences of propagators have a vanishing imaginary
part, which we utilize in the following. The choice of different
(small, positive)  $\eta_i$'s for the propagator would not change our
final result, but is technically more cumbersome.

Using the explicit representation of the momenta in
eq.(\ref{eq:momenta}),
a shift of the loop momenta with a trivial Jacobian according to
\begin{eqnarray}
l_1 & \rightarrow & l_1 + q_z
\nonumber\\
l_0 & \rightarrow & l_0 + l_1
\nonumber\\
k_0 & \rightarrow & k_0 + k_1 \komma
\label{eq:shifts}      
\end{eqnarray}
achieves that the propagators are linear in $l_1$ and $k_1$ and hence
can be written as
\begin{eqnarray}
P_1 & = & (l_0+k_0-e_1)^2+2 (l_0+k_0-e_1)(l_1+k_1)
          -l_\perp^2-k_\perp^2-2 l_\perp k_\perp z-m_1^2+i\eta
\nonumber\\
P_2 & = & (l_0+k_0+e_2)^2+2 (l_0+k_0+e_2)(l_1+k_1)
          -l_\perp^2-k_\perp^2-2 l_\perp k_\perp z-m_2^2+i\eta
\nonumber\\
P_3 & = & (l_0-e_1)^2+2 (l_0-e_1)l_1-l_\perp^2-m_3^2+i\eta
\nonumber\\
P_4 & = & (k_0+e_2+q_z)^2+2 (k_0+e_2+q_z)(k_1-q_z)-k_\perp^2-m_4^2+i\eta
\nonumber\\
P_5 & = & (l_0-q_z)^2+2 (l_0-q_z)(l_1+q_z)-l_\perp^2-m_5^2+i\eta
\nonumber\\
P_6 & = & k_0^2+2 k_0 k_1 -k_\perp^2-m_6^2+i\eta \komma
\end{eqnarray}
where $z$ is the cosine of the angle between $\vec{l}_\perp$ and
$\vec{k}_\perp$.

The volume element in the integral in eq.(\ref{eq:theintegral})
is given by (cf.~\cite{newrepplanar})
\begin{equation}
d^4l \, d^4k = \frac{1}{2} dl_0 \, dk_0 \, dl_1 \, dk_1 \, ds \, dt \, 
               d\alpha \, \frac{dz}{\sqrt{1-z^2}}
\end{equation}
with $s \equiv l_\perp^2$, $t \equiv k_\perp^2$ and the angle $\alpha$
describing the absolute position of $\vec{l}_\perp$ and $\vec{k}_\perp$.
The integration over $\alpha$ gives a trivial factor $2\pi$.

An important difference between the crossed and the planar two-loop
vertex function is that now
necessarily two (instead of one) propagators depend on $z$, 
which, in our notation, are $P_1$ and $P_2$. 
This is a result of the fact that in the
planar case the loop momenta can be arranged to flow through only one
common propagator, which is not possible for the crossed topology.

But after applying a partial fraction decomposition to the integrand
\begin{equation}
\frac{1}{P_1 P_2 P_3 P_4 P_5 P_6} =
    \frac{1}{(P_2-P_1) P_1 P_3 P_4 P_5 P_6}
   +\frac{1}{(P_1-P_2) P_2 P_3 P_4 P_5 P_6}
\label{eq:parfrac1}
\end{equation}
the $z$-dependence in $P_1-P_2$ drops out and the $z$-integration
can be performed as in the planar case:
\begin{equation}
    \int\limits_{-1}^1 \frac{dz}{\sqrt{1-z^2}} \frac{1}{A_k+B_k z+i\eta}
  = \frac{\pi}{\sqrt[c]{A_k^2-B_k^2+2 A_k i\eta}}, \quad k=1,2
\end{equation}
where $P_1=A_1+B_1 z+i\eta$ for the first term in eq.~(\ref{eq:parfrac1})
and $P_2=A_2+B_2 z+i\eta$ in the second, respectively. 
$\sqrt[c]{z}$ denotes the square root of a complex number with a
cut along the positive real axis, whereas $\sqrt{x}$ is the
usual square root of a positive real number (if $x$ has a small imaginary
part of any sign, it can be ignored).

The next step is the integration over $l_1$ and $k_1$ using
Cauchy's theorem for both of them. 
It is important to notice that, as a result of the
partial fraction decomposition in eq.~(\ref{eq:parfrac1}),
in the difference $P_1-P_2$ the
imaginary part of the masses $-i\eta$, which has been 
chosen to be equal for all masses, drops out. 
Therefore not all poles in $l_1$ and $k_1$ lie in the upper or lower
complex half-plane, but some also on the real axis.
The integral can be made meaningful if interpreted as a principal
value integral.\footnote{One can avoid principal value integrals by choosing 
different imaginary parts $\eta_i$ for the propagators, as mentioned
earlier. After having convinced ourselves that the results remain unchanged,
we prefer to follow the route outlined here, for purely technical reasons.} 
Then one has to use the modified Cauchy's theorem 
\cite{henrici1}:
\begin{eqnarray}
\pvint\limits_{-\infty}^{\infty} f(z) \, dz 
 & = & 2 i \pi \sum_i^n \Res(f(z)) \bigg|_{z=z_i, \,\mbox{\scriptsize Im}(z_i) \smalllessgtr 
0}\nonumber\\ 
 & &  + i \pi \sum_j^m \Res(f(z)) \bigg|_{z=z_j, \,\mbox{\scriptsize Im}(z_j)=0},
\end{eqnarray}
where the $z_i$ are the $n$ poles of $f(z)$ inside the closed integration
contour, whereas $z_j$ are $m$ poles along the integration path.
Poles on the path contribute only with half the weight
compared to poles inside the path. For the cases we are confronted
with it is guaranteed that the
function tends to zero sufficiently fast for large $l_1$ and $k_1$, hence
the theorem is applicable.

The integration path has to
be closed either in the upper or in the lower complex half-plane,
depending on the position of the cuts of the square root 
$\sqrt[c]{A^2-B^2+2 A i\eta}$.
Similarly to the planar case, the position
of the cuts is determined by the sign of $l_0+k_0-e_1$ for the
first term in eq.(\ref{eq:parfrac1}) and
by the sign of $l_0+k_0+e_2$ for the second term.

Let us concentrate on the first term of the partial fraction
decomposition, the second can be handled analogously.
Assume $l_0+k_0-e_1<0$, so that we have to close the contour of
the $l_1$ integration in the lower half-plane. Then, some 
propagators have poles inside the path: $P_3$ iff $l_0-e_1>0$
and $P_5$ iff $l_0-q_z>0$. Additionally, $P_2-P_1$ always contributes
with half its residue, since it has a pole on the real axis.
In the following we will call residues, which involve only
$P_3 \dots P_6$, {\em complex} contributions, because all poles
lie in the upper or lower complex half-plane, whereas residues
involving $P_1-P_2$ are called {\em real} contributions, because
the pole lies on the real axis.

For the $k_1$ integration, we close the contour in the same half-plane
as for the $l_1$ integration, depending on the sign of $l_0+k_0-e_1$.
This is not necessary for the $P_2-P_1$
contribution, where the square root becomes independent of $k_1$,
but it is done for consistency.
Let us first assume that we had a $l_1$-pole
in $P_3$. Then we have poles in $k_1$ from $P_4$ iff $k_0+e_2+q_z>0$
and $P_6$ iff $k_0>0$, and a pole in $P_2-P_1$, iff $l_0-e_1<0$.
But the last case is a contradiction to $l_0-e_1>0$ above, so
it does not contribute. 

Analogously, if we had a $l_1$-pole in $P_5$,
we have the same constraints from $P_4$ and $P_6$, but $l_0-q_z<0$
from $P_2-P_1$, which is again a contradiction.

The last case, a $l_1$-pole in $P_2-P_1$, gives contributions from
all other propagators, namely $P_3$ iff $l_0-e_1<0$, $P_4$ iff 
$k_0+e_2+q_z>0$, $P_5$ iff $l_0-q_z<0$ and $P_6$ iff $k_0>0$.

If $l_0+k_0-e_1>0$, the relation operators have to be reversed in
all inequalities.
Out of the {\em complex} combinations, only the pairs $(P_3,P_4)$, 
$(P_5,P_4)$ and
$(P_5,P_6)$ contribute in triangular regions in the $(k_0,l_0)$ plane 
if $l_0+k_0-e_1<0$ (see fig.\ref{fig:complcontr1}). If $l_0+k_0-e_1>0$,
no terms contribute.
All {\em real} residues contribute for $l_0+k_0-e_1<0$ 
as well as for $l_0+k_0-e_1>0$. The areas are not triangles,
but unbound, as can be seen in fig.\ref{fig:realcontr1}.
{
\begin{figure}[ht]
\begin{center}
\fbox{
\epsfig{file=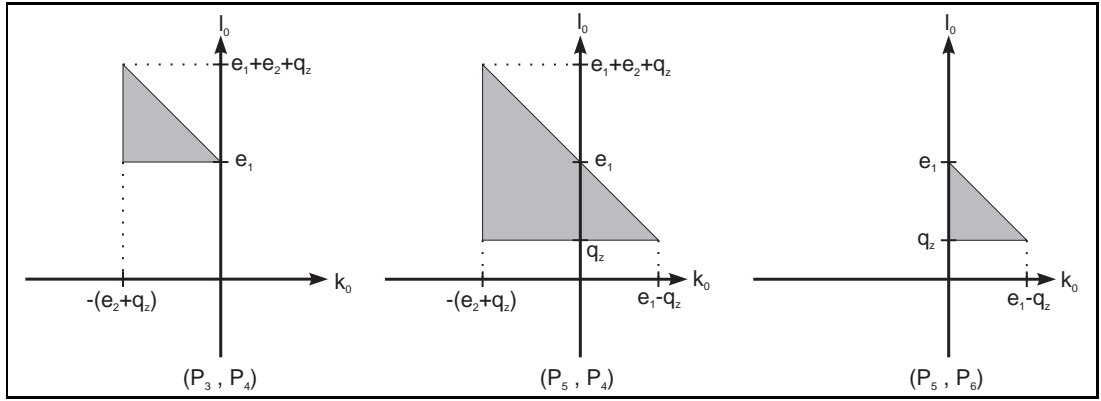,width=14.0cm}
}
\end{center}
\caption {Triangles from the first term of the partial fraction
          decomposition ({\em complex contributions}).}
\label{fig:complcontr1}
\end{figure}
}
{
\begin{figure}[ht]
\begin{center}
\fbox{
\epsfig{file=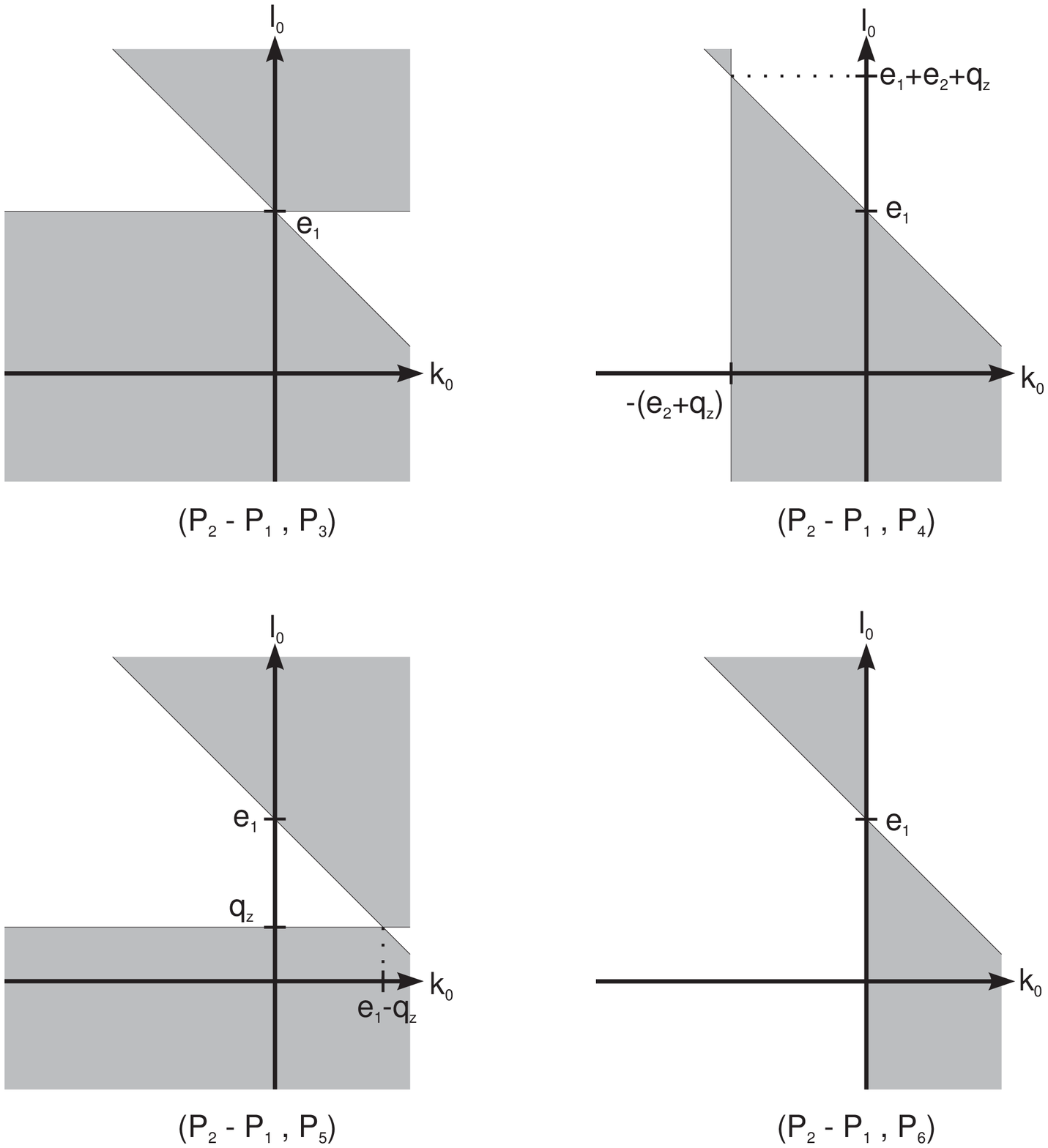,width=10.0cm}
}
\end{center}
\caption {Unbound areas from the first term of the partial fraction
          decomposition ({\em real contributions}).}
\label{fig:realcontr1}
\end{figure}
}

The second term of the partial fraction decomposition also gives us
three contributing {\em complex} residues (all for $l_0+k_0+e_2>0$,
fig.\ref{fig:complcontr2})
and four {\em real} residues, which have the same constraints
as for the 
first term, but with $l_0+k_0-e_1 \lessgtr 0$ replaced by 
$l_0+k_0+e_2 \lessgtr 0$ (fig.\ref{fig:realcontr2}).
{
\begin{figure}[ht]
\begin{center}
\fbox{
\epsfig{file=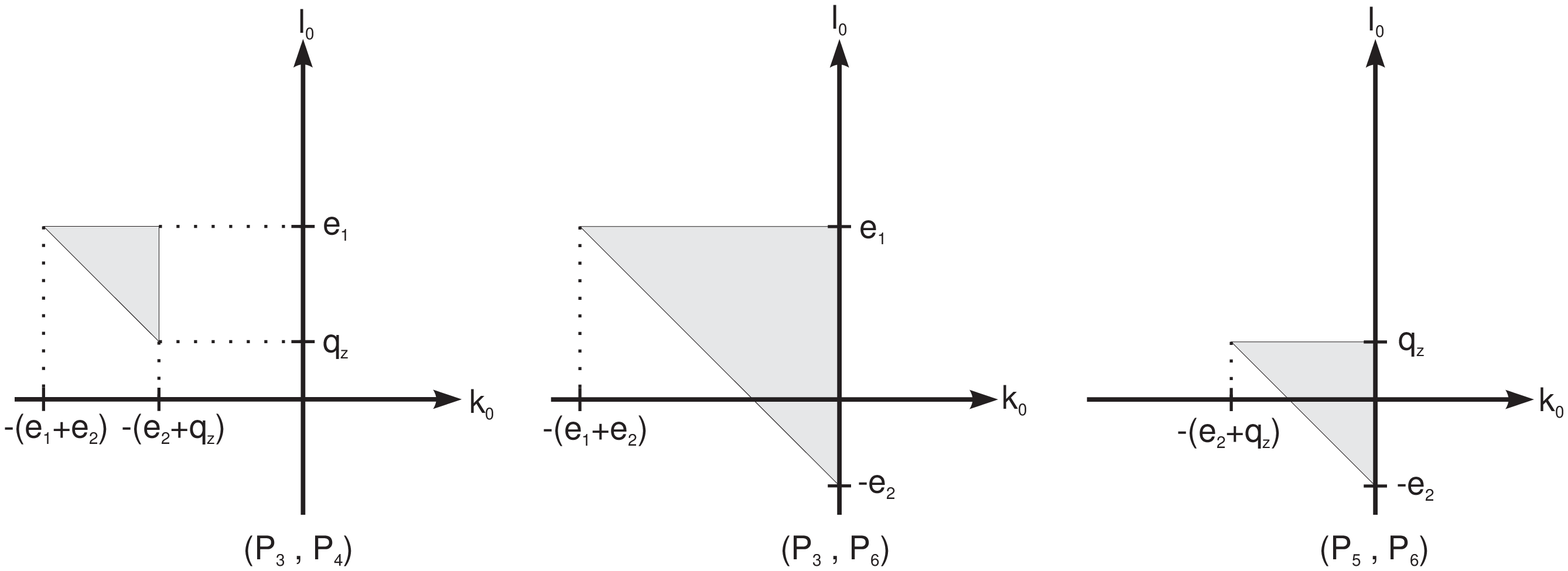,width=14.0cm}
}
\end{center}
\caption {Triangles from the second term of the partial fraction
          decomposition ({\em complex contributions}).}
\label{fig:complcontr2}
\end{figure}
}
{
\begin{figure}[ht]
\begin{center}
\fbox{
\epsfig{file=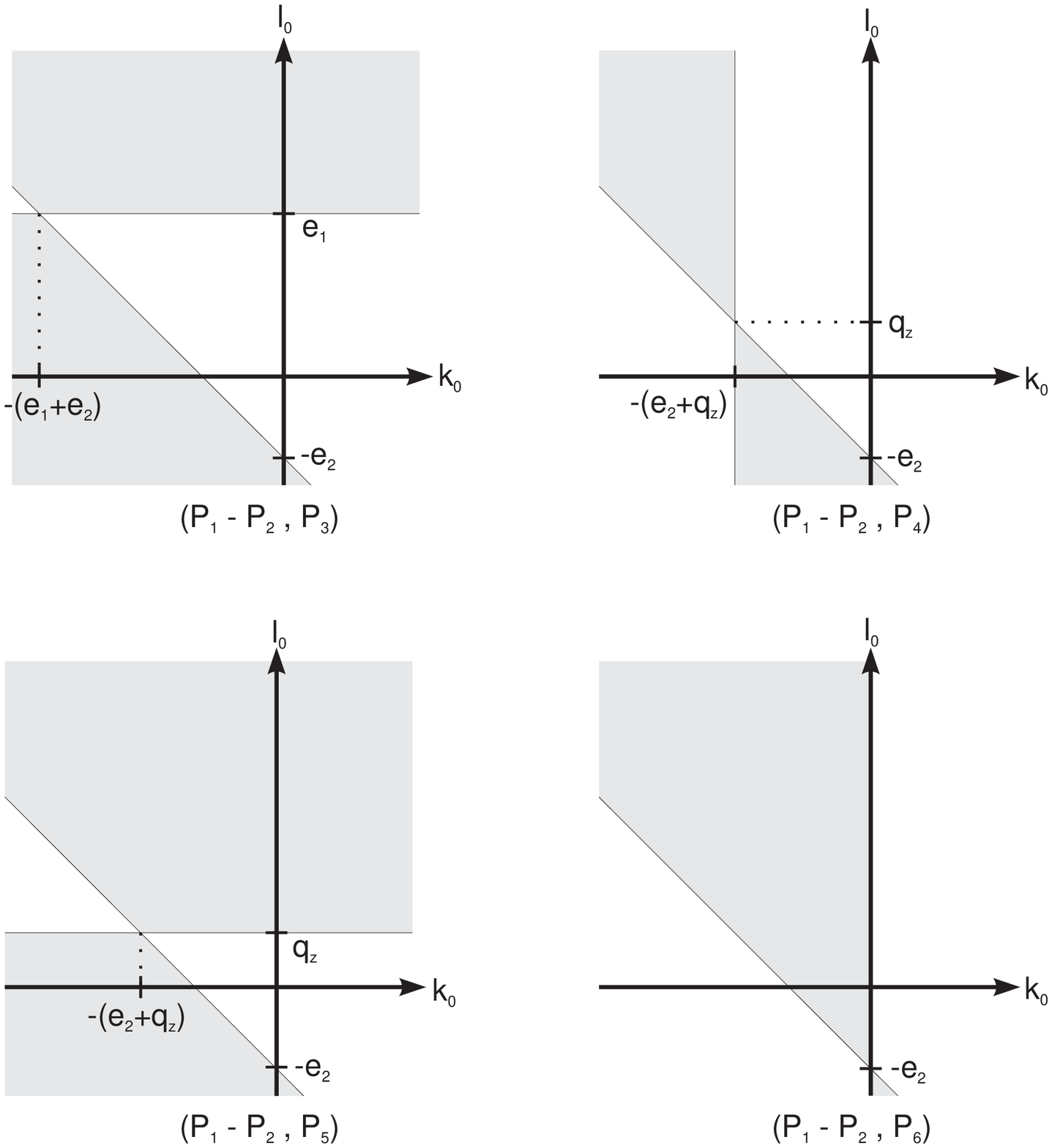,width=10.0cm}
}
\end{center}
\caption {Unbound areas from the second term of the partial fraction
          decomposition ({\em real contributions}).}
\label{fig:realcontr2}
\end{figure}
}

If we compare the {\em real} contributions from both terms of the partial
fraction decomposition eq.(\ref{eq:parfrac1}), we notice that
the terms for two corresponding residues
(e.g. $(P_2-P_1, P_3)$ and $(P_1-P_2,P_3)$)
are equal, except for an overall minus sign. As a consequence
they cancel in the area where both contribute together.
This is everywhere the case except in the strip $-e_2 < l_0+k_0 < e_1$
(see fig.\ref{fig:realcontr3}).
Furthermore it can be shown that all four {\em real} contributions
from fig.\ref{fig:realcontr3} add up to zero where they all 
contribute together. This is the case everywhere outside the 
finite area
shown in fig.\ref{fig:realcontr4}, which is also the joined area of
all triangles in figs.\ref{fig:complcontr1}
and \ref{fig:complcontr2} from the {\em complex} residues.
{
\begin{figure}[ht]
\begin{center}
\fbox{
\epsfig{file=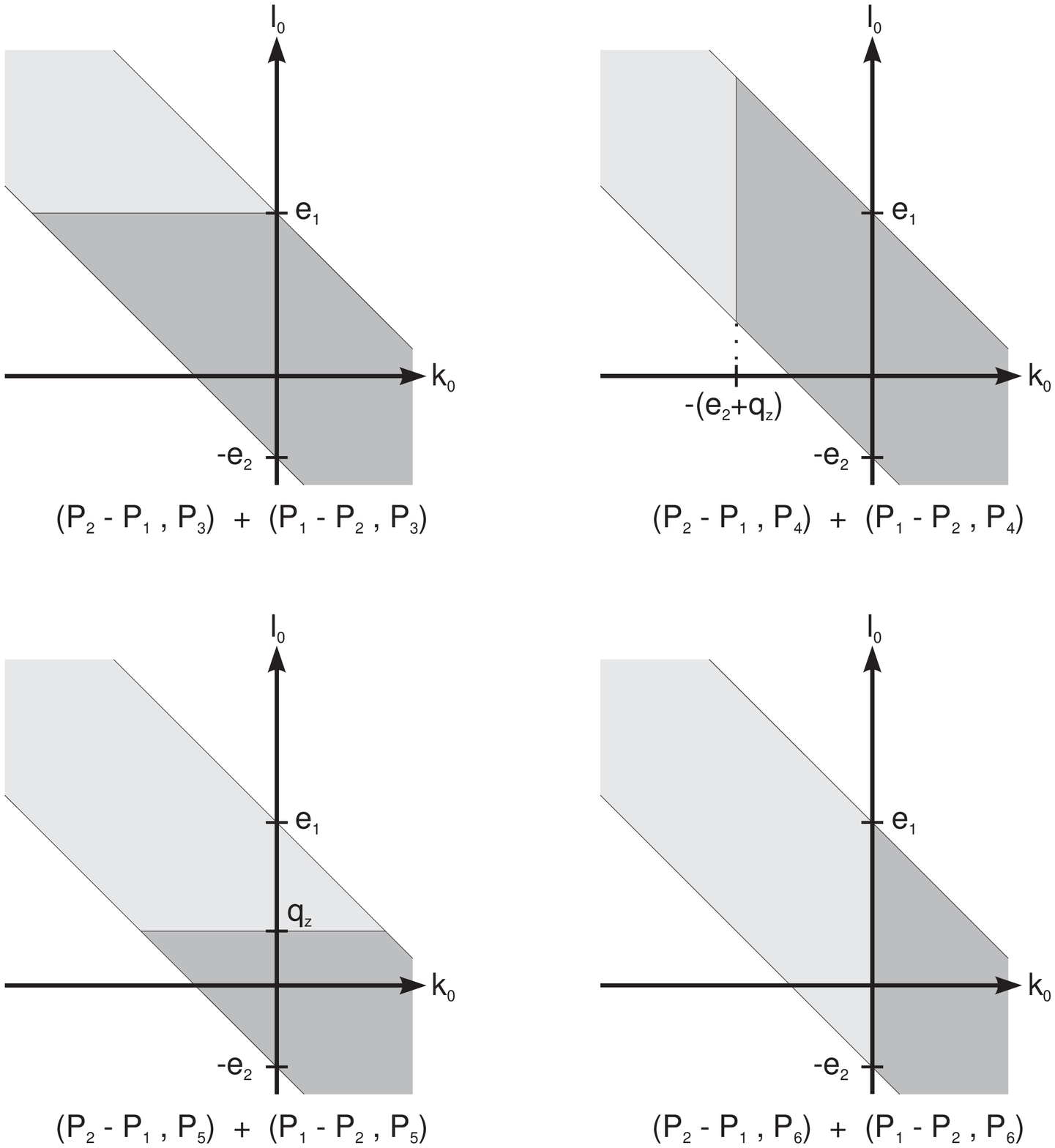,width=10.0cm}
}
\end{center}
\caption {Sum of both terms of the pfd for the {\em real} contributions
(dark shading from the first, light shading from the second term $=$
$-$ first 
term.)}
\label{fig:realcontr3}
\end{figure}
}
{
\begin{figure}[ht]
\begin{center}
\fbox{
\epsfig{file=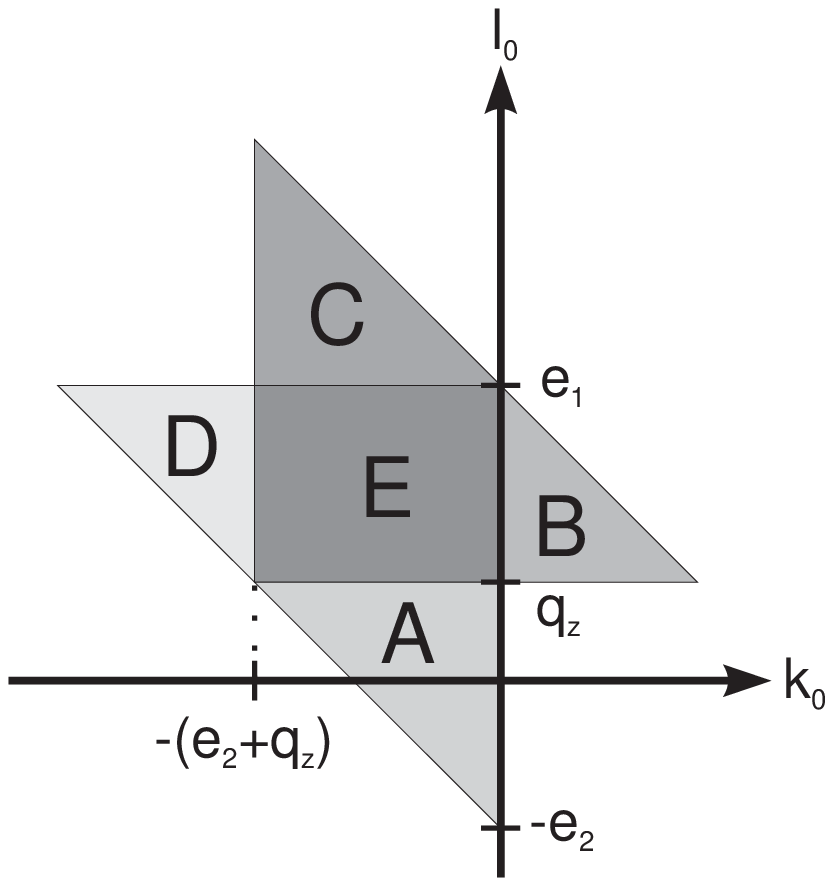,width=5.5cm}
}
\end{center}
\caption {Sum of all four terms from fig.~\ref{fig:realcontr3}.}
\label{fig:realcontr4}
\end{figure}
}

After the residue integrations we end up with a four-fold integral
representation which is similar in nature, but slightly more 
complicated in its technical appearance than  the
planar case. It can be written in the general form
\begin{eqnarray}
V(q_1,q_2) & = & \sum_j \iint\limits_{A_j} \!\! dl_0 \, dk_0 
   \int\limits_0^\infty\!\! ds
   \int\limits_0^\infty\!\! dt \: C(k_0,l_0) 
\nonumber\\
 & & \hspace{-1.0cm}
     \frac{1}{\tila_1 t+\tilb_1+\tilc_1 s}
     \frac{1}{\tila_2 t+\tilb_2+\tilc_2 s}
     \frac{1}{\tila_3 t+\tilb_3+\tilc_3 s}
     \frac{1}{\sqrt[c]{(a t+b+c s)^2-4 s t}} \pkt
\label{eq:4fold}
\end{eqnarray}
The $j$-sum runs over all areas $A_j$ depicted in figs.\ref{fig:complcontr1}
-- \ref{fig:realcontr2}. The coefficients $C(k_0,l_0)$, $\tila_i$, $\tilb_i$,
$\tilc_i$ ($i=1 \dots 3$), $a$, $b$ and $c$ are in general dependent on $j$,
$l_0$ and $k_0$. $\tilb_i$ and $b$ have an infinitesimal imaginary part, 
all other
coefficients are real. Some coefficients are vanishing: for the
{\em complex} residues, $\tila_1$ and $\tilc_3$ are zero, and we
have one pure pole in $s$, one in $t$ and a mixed $(s,t)$ pole.
For the {\em real} residues, either $\tilc_3=0$ or $\tila_3=0$, i.e. we have
two mixed poles and either one pure pole in $t$ or in $s$.
In the latter case we can relabel $s {\displaystyle\leftrightarrow} t$,
together
with exchanging $\tila_i \leftrightarrow \tilc_i$ and $a \leftrightarrow c$,
so that we always have exactly two poles in $s$, either pure or mixed,
and $c_3=0$. In the planar case we always had one pure $s$ pole and
two pure $t$ poles, with mixed poles altogether absent.

To proceed further with the $s$ and $t$ integrations, we apply a partial
fraction decomposition in $s$, followed by a similar
partial fraction decomposition in $t$.
\begin{eqnarray}
\lefteqn{
\frac{1}{\tila_1 t+\tilb_1+\tilc_1 s}
\frac{1}{\tila_2 t+\tilb_2+\tilc_2 s}
\frac{1}{\tila_3 t+\tilb_3}
} & &
\nonumber\\
& = & C'
\left(
\frac{1}{t+\bar{t}_{02}} -
\frac{1}{t+\bar{t}_{01}}
\right)
\left(
\frac{1}{s+\bar{s}_{02}(t)} -
\frac{1}{s+\bar{s}_{01}(t)}
\right)
\end{eqnarray}
with 
\begin{eqnarray}
C' & = & \frac{1}{\tilc_2(\tila_1\tilb_3-\tilb_1\tilc_3)-
                  \tilc_1(\tila_2\tilb_3-\tilb_2\tilc_3)}
\nonumber\\
\bar{t}_{01} & = & \frac{\tilb_3}{\tilc_3}
\nonumber\\
\bar{t}_{02} & = & \frac{\tilc_2\tilb_1-\tilc_1\tilb_2}
                        {\tilc_2\tila_1-\tilc_1\tila_2} \pkt
\nonumber\\
\bar{s}_{01}(t) & = & \frac{\tila_1 t+\tilb_1}{\tilc_1}
\nonumber\\
\bar{s}_{02}(t) & = & \frac{\tila_2 t+\tilb_2}{\tilc_2} \pkt
\end{eqnarray}
Thus we have to calculate four integrals of the form
\begin{equation}
V' = \int\limits_0^\infty \frac{dt}{t+\bar{t}_0} 
     \int\limits_0^\infty \frac{ds}{s+\bar{s}_0(t)}
\frac{1}{\sqrt[c]{(a t+b+c s)^2 - 4st}}
\end{equation}
with $s_0(t)$ being either a linear function of $t$ or a constant.
Now we have to split these into real and imaginary part. 
It can be shown that $b$ always has a positive small imaginary part,
whereas the sign of the imaginary part
of $\bar{s}_0$ and $\bar{t}_0$ is a function of $l_0$ and $k_0$, so that in
general $\bar{s}_0=s_0 \pm i\eta$ and 
$\bar{t}_0=t_0 \pm i\eta$. In contrast, in 
the planar case the imaginary part was always
negative. 

Now there are three possible sources for an imaginary part of $V'$: either
$s_0$, $t_0$ or the argument of the square root can become
negative. In the first two cases the imaginary part can be extracted
using
\begin{equation}
\label{eq:pvplusipidelta}
\lim_{\eta \to 0} \int\limits_0^\infty \frac{dx}{x+x_0 \pm i\eta} f(x)
 = \pvint\limits_0^\infty \frac{dx}{x+x_0} f(x)
   \mp i \pi \int\limits_0^\infty \delta(x+x_0) f(x) dx \pkt
\end{equation}
To analyze the contribution from the square root, we have to
distinguish between {\em complex} and {\em real} residues.
For the {\em complex} case, the area in the $s-t$ plane, where the argument
is negative, is an ellipse. For a given positive $t$, the $s$ values on the
border of the ellipse can be calculated as
\begin{equation}
\label{eq:sigmas}
\sigma_{1/2} = 
    \frac{1}{c^2}\left[ \sqrt{t} \pm \sqrt{t(1-a c)-b c} \right] ^2 \komma
\end{equation}
which has real and positive solutions for $0 < t < bc/(1-ac)$ if $b>0$, since
$1-ac<0$, cf.~fig.~\ref{fig:ellipse}. 
For the {\em real} residues, $a=-1$ and $c=-1$,
therefore the area where the argument becomes negative
is unbounded with a parabola as its boundary (fig.~\ref{fig:parabel}).
{
\begin{figure}[ht]
\begin{center}
\fbox{
\epsfig{file=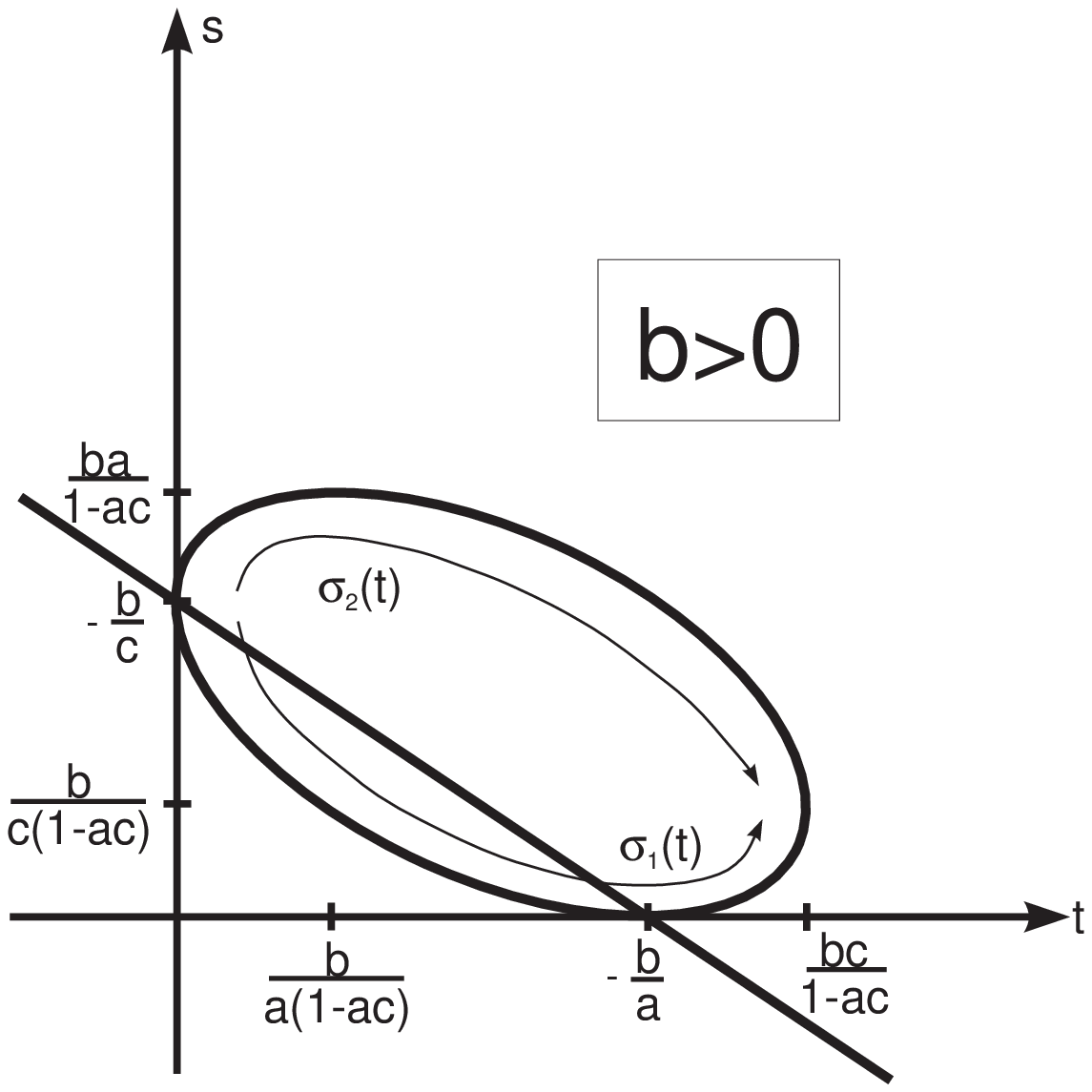,width=5.5cm}
}
\end{center}
\caption {The ellipse.}
\label{fig:ellipse}
\end{figure}
}
{
\begin{figure}[ht]
\begin{center}
\fbox{
\epsfig{file=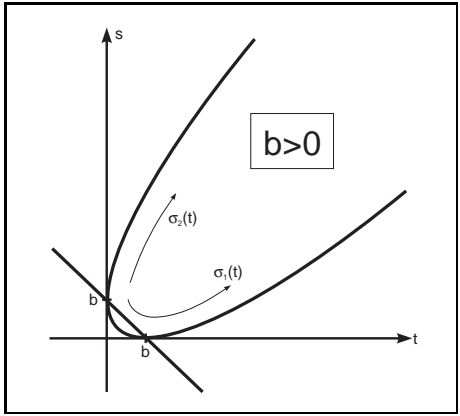,width=5.5cm}
}
\end{center}
\caption {The parabola.}
\label{fig:parabel}
\end{figure}
}

Now we can do the $s$-integration with the aid of an Euler's change
of variables \cite{fichtenholz} and obtain
\begin{equation}
V' = \frac{1}{|c|}\int\limits_0^\infty \frac{1}{t+t_0}
\left( f(s_0,\sigma_i) + i k(t) g(s_0,\sigma_i) \right) \, dt
\end{equation}
with 
\begin{equation}
\label{eq:sint1}
f(s_0,\sigma_i) = \left\{
\begin{array}{lll}
  \frac{2}{\sqrt{-R}}
  \arctan\left(\frac{s_0\pm\wsigsig}{\sqrt{-R}}\right)
  & \mbox{if} \quad \sigma_1 < -s_0 < \sigma_2
  & \mbox{(i)}
\\
  -\frac{1}{\sqrt{R}}
  \log\left|\frac{s_0\pm\wsigsig+\sqrt{R}}
                 {s_0\pm\wsigsig-\sqrt{R}}
  \right|
  & \mbox{else}
  & \mbox{(ii)}
\end{array}\right.
\end{equation}
and
\begin{equation}
\label{eq:sint2}
g(s_0,\sigma_i) = \left\{
\begin{array}{lll}
  \frac{i}{\sqrt{-R}}
  & \mbox{if} \quad \sigma_1 < -s_0 < \sigma_2
  & \mbox{(i)}
\\
  \frac{1}{\sqrt{R}}
  & \mbox{else}
  & \mbox{(ii)}
\end{array}\right.
\end{equation}
where we
note that $s_0$ and $\sigma_i$ are functions of $t$ and
further  $R=\wquad$
and $k=-2,-1,0,1$ or 2, depending on the relative magnitudes of
$s_0$ and $\sigma_{1/2}$ and the sign of the imaginary part of
$s_0$. One has to choose $+\wsigsig$ in eqs.(\ref{eq:sint1}) and 
(\ref{eq:sint2}) if $t>-b/a$, and $-\wsigsig$ else.
The condition $\sigma_1 < -s_0 < \sigma_2$ is equivalent to $R<0$.

The $t$-intervals where $k$ is constant, together with the corresponding $k$
values,
can be calculated by solving $s_0=0$ and $s_0=-\sigma_i$ for $t$. 
With another Euler's change of variables,
using eq.(\ref{eq:pvplusipidelta}) and exploiting the relevant
 properties of the
functions $\log(x)$ and $\arctan(x)$, the $t$-integration leads to
expressions of the form
\begin{equation}
\pvint\limits_{x_1}^{x_2} \frac{1}{x^2+px+q}
\quad\mbox{from} \;\; g(s_0,\sigma_i) \komma
\end{equation}
\begin{equation}
\pvint\limits_{x_1}^{x_2} \frac{\arctan(rx+s)}{x^2+px+q}
\quad\mbox{from}\; f(s_0,\sigma_i),\; \mbox{case (i)}
\end{equation}
and
\begin{equation}
\pvint\limits_{x_1}^{x_2} \frac{\log|x^2+rx+s|}{x^2+px+q}
\quad\mbox{from}\; f(s_0,\sigma_i),\; \mbox{case (ii)} \komma
\end{equation}
which can be expressed in terms of logarithms, arcus-tangens, dilogarithms
and Clausen's functions \cite{lewin}, as it was the case
for the planar topology.
The full result is, as expected, a lengthy expression in terms
of these special functions, which we cannot list here. 
We rather follow the philosophy
of \cite{newrepplanar} and present examples in the next section.

We stress that to obtain stable numerical results 
in the two dimensional integral over $l_0$ and $k_0$
it is very important to
add all contributions ({\em real} and {\em complex}) to
the same $l_0$ and $k_0$, because eq.(\ref{eq:parfrac1})
introduces some artificial divergences which cancel in the sum.

As expected, the numerical integrand for the crossed topology is
of the same nature as for the planar case, but involves more
terms and different cases. This naturally increases the amount
of CPU time needed to obtain the requested accuracy:
as a thumb rule, we found that the crossed topology demands
5-10 times more time compared to the planar case.

The threshold behaviour can be examined with the four-fold integral
representation eq.(\ref{eq:4fold}). 
The crossed vertex function has three two-particle thresholds and 
six three-particle thresholds. 
As stated above a possible source for an imaginary part is a
negative argument of the square root. A necessary condition for
this to happen is $b>0$ for some $l_0$, $k_0$ inside the integration
region. It can be shown that each three-particle
threshold corresponds to one of the six $b$ coefficients of 
the {\em complex} triangles \cite{xloops}. 
The other $b$ coefficients belonging
to the {\em real} regions are identical and correspond to the
two-particle threshold $q^2>(m_1+m_2)^2$.
The remaining two-particle thresholds $q_1^2>(m_3+m_5)^2$ and
$q_2^2>(m_4+m_6)^2$ can be identified with the coefficients
$\bar{s}_{01}$ (when $\tila_1 = 0$) and $\bar{t}_{01}$ respectively
which produce an imaginary part when $\bar{s}_{01}$ and $\bar{t}_{01}$
become negative.
A possible four-particle threshold $q^2>(m_3+m_4+m_5+m_6)^2$ disconnects
the graph into three parts and is therefore a combination of the two
two-particle thresholds $q_1^2>(m_3+m_5)^2$ and $q_2^2>(m_4+m_6)^2$.

\section{Examples}

\begin{figure}[ht]
\begin{center}
\fbox{
\epsfig{file=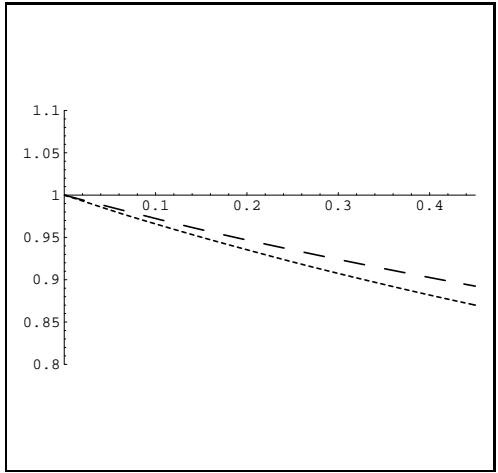,width=6cm}
}
\end{center}
\caption {Ratio of crossed vertex-function to master two-point function
for $q_1$ (short dash) and $q_2 \to 0$ (long dash).}
\label{fig:exmaster}
\end{figure}

Only few analytical and/or numerical results are known for
the crossed vertex function to compare with.
Besides the symmetries with respect to internal masses and external
momenta, some limiting cases can be checked.
In the case of zero momentum transfer
the crossed vertex function reduces to the
master two-point topology \cite{dirk1,baub} with a squared propagator.
\fig{exmaster} shows these limits for some arbitrary masses 
($m_1=2$, $m_2=3$, 
$m_3=1$, $m_4=1.5$, $m_5=0.6$ and $m_6=0.2$),
$q^2=1$ and either $q_1 \to 0$ or $q_2 \to 0$.
Additionally, for vanishing $q^2$ we obtain the vacuum bubble
calculated in \cite{davydtausk}.

\begin{figure}[ht]
\begin{center}
\fbox{
\epsfig{file=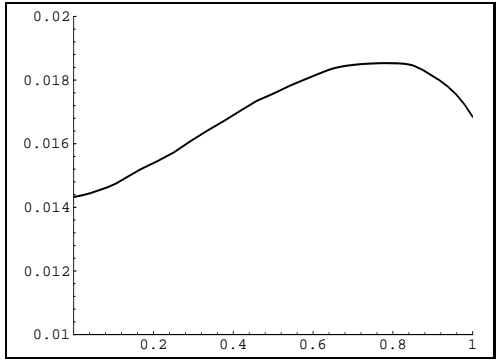,width=6cm}
}
\end{center}
\caption {Limit of all masses vanishing.}
\label{fig:zeromass}
\end{figure}

For the case of all internal masses vanishing an analytical formula
in known \cite{davyd2}. This limit is approached smoothly in \fig{zeromass}
for all masses going to zero simultaneously and 
$e_1=4$, $e_2=3$ and $q_z=1$.

\begin{figure}[ht]
\begin{center}
\fbox{
\epsfig{file=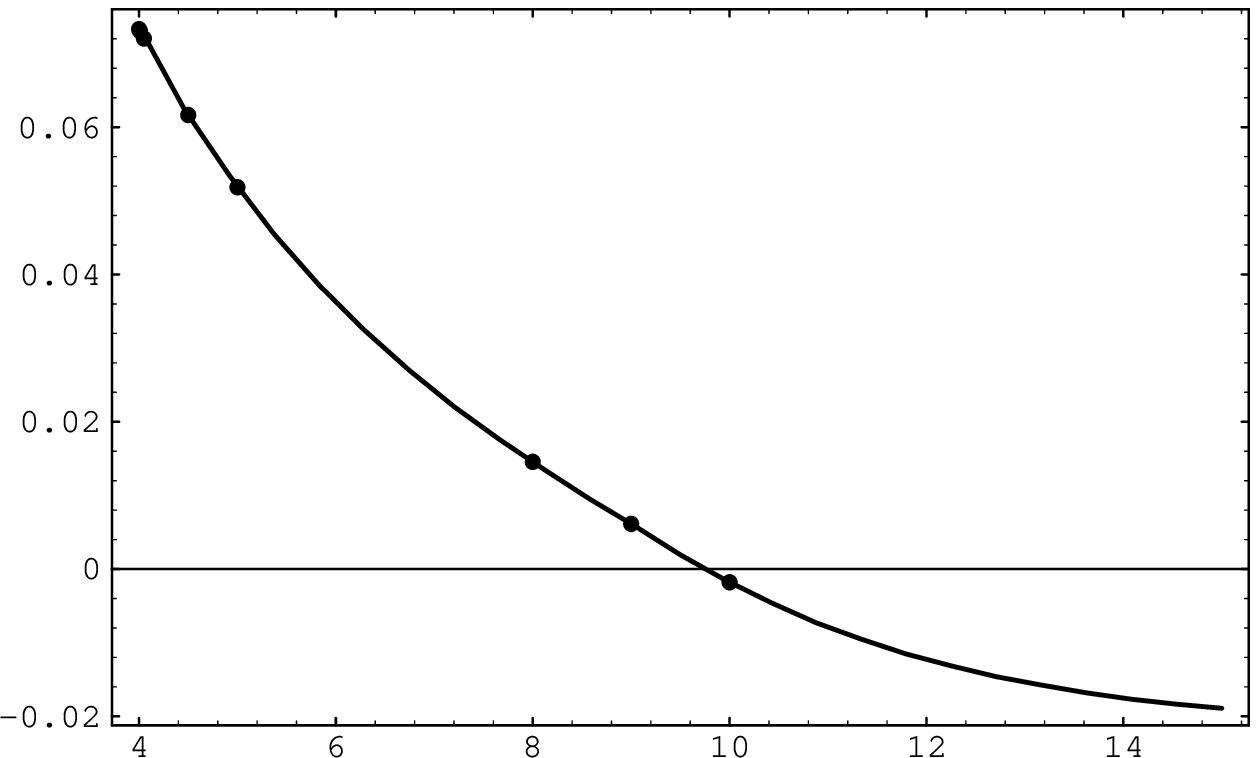,width=5.5cm}
}
\fbox{
\epsfig{file=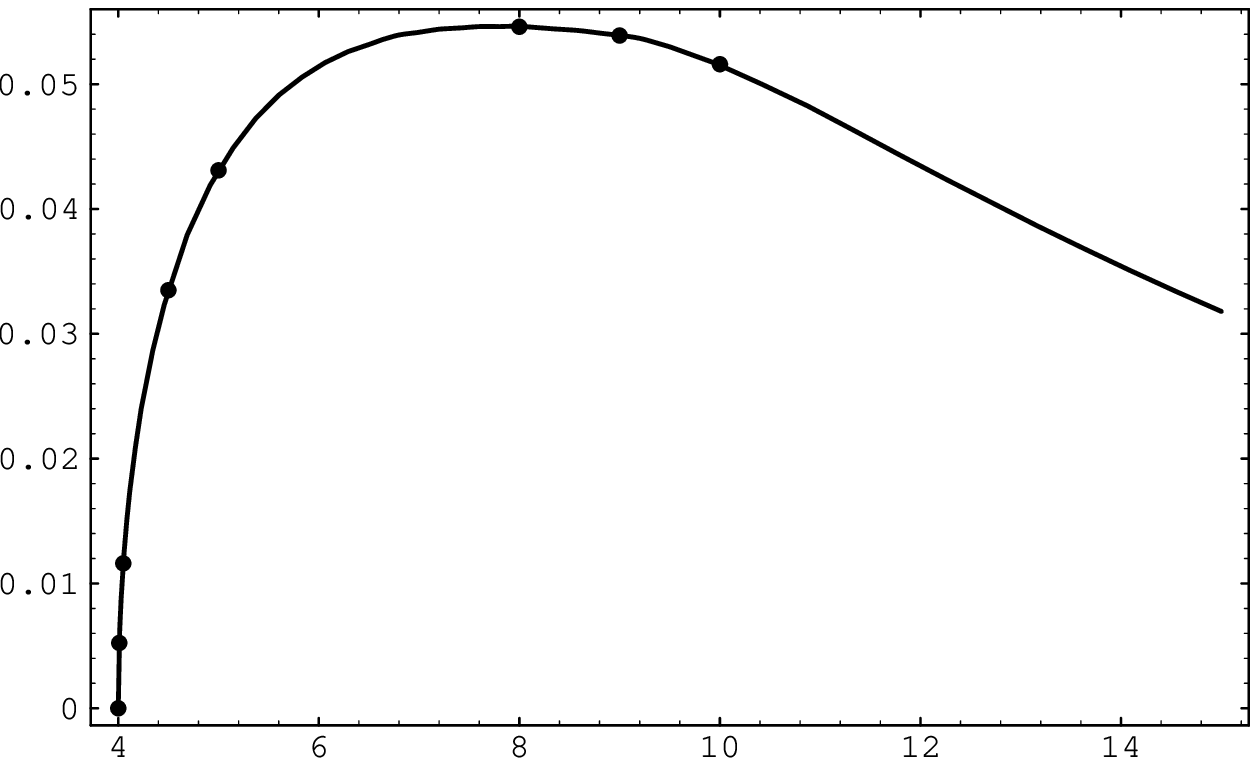,width=5.5cm}
}
\end{center}
\caption {All masses equal, $q_1^2=q_2^2=0$, real part (left) and 
imaginary part (right).}
\label{fig:tarasov}
\end{figure}

In the limit of the momenta of the outgoing particles on the light-cone
($q_1^2=q_2^2=0$) and all internal masses equal this diagram can
be calculated with the small momentum expansion technique \cite{tarasov}.
A comparison for real and imaginary part is show in \fig{tarasov}.
We can obtain this limit easily by transforming our parallel space
coordinates to light cone coordinates, which does not interfere
with our subsequent steps.

\begin{figure}[ht]
\begin{center}
\fbox{
\epsfig{file=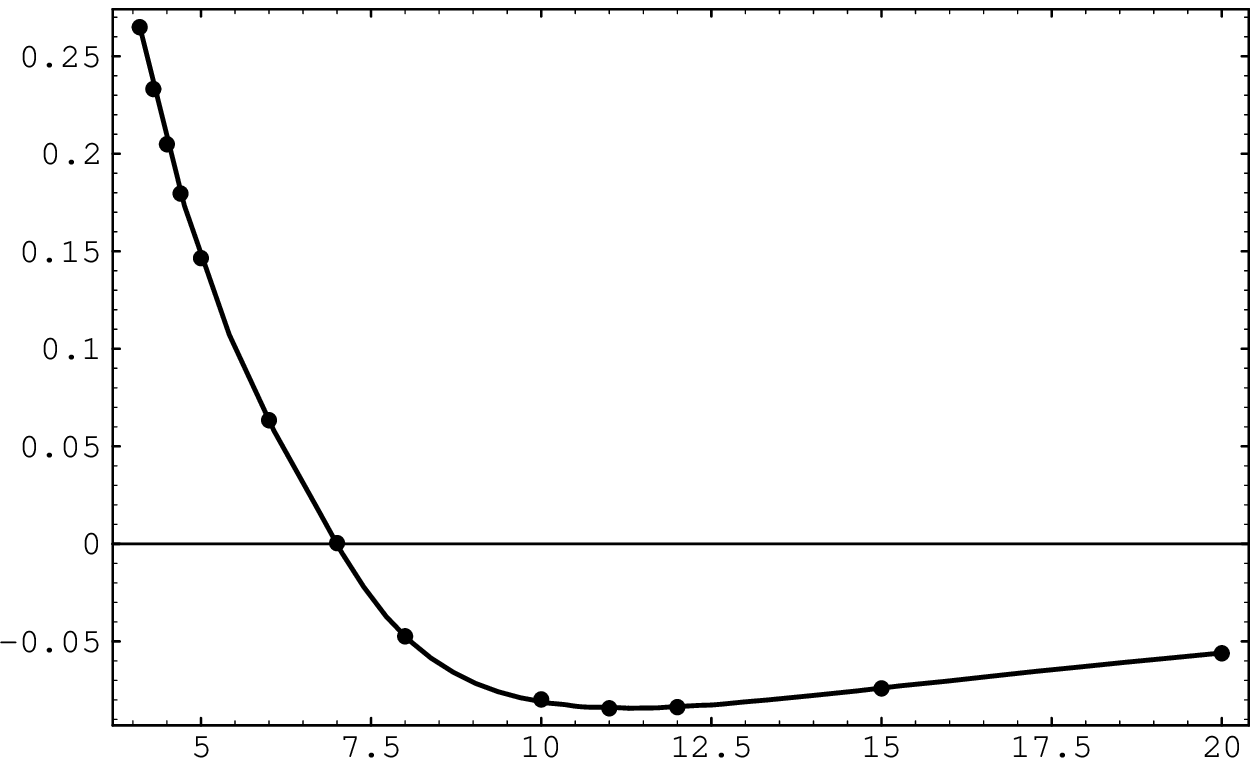,width=6cm}
}
\end{center}
\caption {Decay $\rm Z \to t\bar{t}$, real part.}
\label{fig:kato}
\end{figure}

As a last example in \fig{kato} we show the decay $\rm Z \to t \bar{t}$
with the exchange of two Z bosons.
This diagram has been calculated in \cite{kato} by
a five-dimensional numerical integration over the Feynman parameters. 

\section*{Conclusions}
In this paper, we demonstrated the calculation of the two-loop
vertex function for the crossed topology. 
We outlined how to achieve the reduction to a manageable
two-fold integral representation,
and verified its correctness by comparison with the
literature, wherever data were available.
Further, our integral representation was used
in \cite{FKKR}, with results in perfect agreement
with the expectations.

Our results complement the
results for the planar topology in \cite{newrepplanar}.
Together, the scalar two-loop three-point function is now available in 
$D=4$ dimensions for all topologies, all masses, and arbitrary
external momenta. 
Some degenerated topologies were already given in
\cite{FKKR}, obtained by similar methods. Such degenerated cases typically
also appear when one confronts tensor integrals. For the future,
we plan to incorporate such cases in the package XLOOPS \cite{xloops}.
Also, code which implements the results presented here will
be incorporated there.

To our knowledge, no other method is at this stage able to deliver
reliable results for the massive two-loop vertex function in such
generality and accuracy.
 
\section*{Acknowledgements}
We like to thank David Broadhurst, Jochem Fleischer, Bernd Kniehl, 
Kurt Riesselmann,
Karl Schilcher and Volodya Smirnov
for interesting discussions and support,
and the participants and organizers of AIHENP96 (Lausanne, September 1996)
for a stimulating workshop.
D.K.~thanks the DFG for support. 
This work was supported in part by HUCAM grant CHRX-CT94-0579.

\end{document}